\documentclass[sigconf,dvipsnames,nonacm]{acmart}

\AtBeginDocument{%
  }

\setcopyright{acmlicensed}
\copyrightyear{2024}
\acmYear{2024}
\acmDOI{XXXXXXX.XXXXXXX}

\acmConference[CIKM '24]{Make sure to enter the correct
  conference title from your rights confirmation email}{October 03--05,
  2024}{Boise, ID}
\acmISBN{978-1-4503-XXXX-X/18/06}

\usepackage{algorithm}
\usepackage{algpseudocode}

\usepackage[disable]{todonotes}
\usepackage{cleveref}
\usepackage{natbib}
\usepackage{amsmath}
\usepackage{subcaption}
\usepackage{soul}

\newcommand{\jccomment}[1]{\todo[color=orange!20, inline]{\textbf{Joe:} #1}}
\newcommand{\yycomment}[1]{\todo[color=red!20, inline]{\textbf{Yixiang:} #1}}

\newcommand{\enc}[1]{\llbracket #1 \rrbracket}
\newcommand{\inv}{\operatorname{\texttt{Inv}}}
\newcommand{\eeq}{\operatorname{\texttt{EEq}}}
\newcommand{\comp}{\operatorname{\texttt{Comp}}}

\algnewcommand{\LineComment}[1]{\State \(\triangleright\) #1}

\begin{document}

\title{Feasibility of Privacy-Preserving Entity Resolution on Confidential Healthcare Datasets Using Homomorphic Encryption}

\author{Yixiang Yao}
\email{yixiangy@usc.edu}
\affiliation{%
  \institution{University of Southern California}
  \city{Los Angeles}
  \state{California}
  \country{USA}
}

\author{Joseph Cecil}
\email{jcecil@isi.edu}
\affiliation{%
  \institution{Information Sciences Institute}
  \city{Waltham}
  \state{Massachusetts}
  \country{USA}
}

\author{Praveen Angyan}
\email{praveen.angyan@med.usc.edu}
\affiliation{%
  \institution{Southern California Clinical and Translational Science Institute}
  \city{Los Angeles}
  \state{California}
  \country{USA}
}

\author{Neil Bahroos}
\email{neil.bahroos@med.usc.edu}
\affiliation{%
  \institution{Southern California Clinical and Translational Science Institute}
  \city{Los Angeles}
  \state{California}
  \country{USA}
}

\author{Srivatsan Ravi}
\email{srivatsr@usc.edu}
\affiliation{%
  \institution{University of Southern California}
  \city{Los Angeles}
  \state{California}
  \country{USA}
}


\begin{abstract}

Patient datasets contain confidential information which is protected by laws and regulations such as HIPAA and GDPR. Ensuring comprehensive patient information necessitates privacy-preserving entity resolution (PPER), which identifies identical patient entities across multiple databases from different healthcare organizations while maintaining data privacy. Existing methods often lack cryptographic security or are computationally impractical for real-world datasets. We introduce a PPER pipeline based on AMPPERE, a secure abstract computation model utilizing cryptographic tools like homomorphic encryption. Our tailored approach incorporates extensive parallelization techniques and optimal parameters specifically for patient datasets. Experimental results demonstrate the proposed method's effectiveness in terms of accuracy and efficiency compared to various baselines.

\end{abstract}



\maketitle

\section{Introduction}

Patients visit different healthcare facilities to receive primary care, specialized treatments such as obstetrics or dentistry, urgent or emergency services, or to fill prescriptions. However, optimal and safe healthcare often necessitates a unified medical history for each patient. Moreover, healthcare research is frequently multidisciplinary, necessitating the merging of datasets from various research teams. During health crises like pandemics, it is crucial for organizations to share health datasets swiftly and securely. An example is the National COVID Cohort Collaborative (N3C), which was established to facilitate the sharing of secure and de-identified clinical data across the US for COVID-19 research.
\textit{Entity resolution (ER)} is the task of finding records that refer to the same entity across different data sources~\cite{papadakis2020blocking}. This technique aligns with the purpose of identifying the same patients or research participants to consolidate data about each individual in multidisciplinary and collaborative research.

When applied to patient datasets, ER faces significant challenges due to privacy regulations such as the Health Insurance Portability and Accountability Act (HIPAA) in the United States and the General Data Protection Regulation (GDPR) in the European Union. 
The Privacy Rule provided by the HIPAA~\cite{hipaa-privacy-rule} gives patients rights over their health information and sets rules and limits on who can look at and receive their health information. HIPAA specifies 18 elements in health data~\cite{hipaa-18-elements} that are considered identifiers.
Additionally, certain populations in healthcare are at particularly high risk if their private information is exposed, such as individuals with mental disabilities or mental health conditions, HIV, or those experiencing homelessness. In healthcare and research, merging datasets by sharing the full dataset is inherently risky. Even without explicit identifiers, medical data containing rare conditions can be sufficient to uniquely identify a patient.

Existing solutions to address these challenges include the use of \textit{privacy-preserving entity resolution (PPER)} techniques, which enable the matching of records across datasets while keeping the data encoded and ensuring that individual privacy is maintained \cite{vatsalan2013taxonomy,vatsalan2017privacy,gkoulalas2021modern}. These techniques encompass methods such as hashing, masking, perturbation, and probabilistic approaches, as well as the use of natural language models to encode records into multi-dimensional embedding vectors. However, these methods are either susceptible to various attacks or require a trade-off between data utility and privacy. 
As a solution, AMPPERE~\cite{yao2021amppere}, as an platform-agnostic PPER framework leveraging either secure multi-party computation or fully homomorphic encryption, is proposed to guarantee the cryptographic security of the data without sacrificing utility. However, as a universal abstract model, AMPPERE encounters efficiency challenges when scaling to large practical datasets, as it requires compatibility with heterogeneous underlying secure multi-party computation protocols or homomorphic encryption schemes.

In this work, we focus on resolving the real-world PPER challenge of healthcare datasets, aiming to ensure HIPAA compliance and minimize the risk of exposing private information. 
We develop our open-source solution using AMPPERE as the foundation, \ul{ensuring precise computational accuracy and provable privacy}.
We extensively parallelize the execution by employing various techniques and optimize the parameters specifically for the datasets. Subsequently, we conduct comprehensive experiments comparing our optimized method with baselines, as well as ablation studies, to verify the effectiveness of each optimized component.

\yycomment{citations for medical related-stuff}

\begin{figure*}[ht!]
    \centering
    \includegraphics[width=1\linewidth]{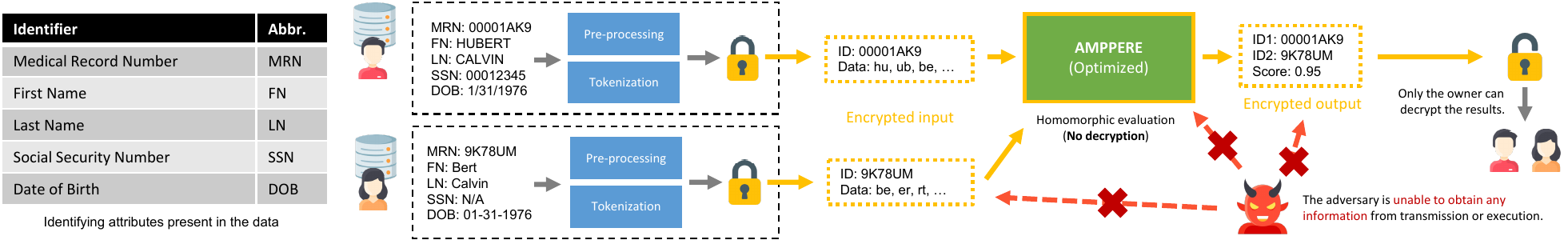}
    \caption{The overview of the dataset and the pipeline. Arrows and boxes in \textcolor{Dandelion}{orange} indicate the data is in cipher. Each record, after pre-processing and tokenization, is encrypted into ciphertext. All records from both datasets are then sent to the optimized AMPPERE pipeline, which homomorphically evaluates the potential matches and produces record pairs with scores in encrypted form. Finally, the data owners decrypt the results collectively, whereas the adversary cannot gather any information.}
    \label{fig:overview}
\end{figure*}

\section{Patient Mortality Datasets}

There are numerous PPER tasks in the healthcare domain. In this study, we focus on one specific task related to patient mortality.
Patients in a healthcare organization are only marked as deceased if they pass away during care. However, research into the causes of mortality necessitates up-to-date information. Electronic Medical Records (EMR) systems are rarely updated with the mortality statuses of former patients. Acquiring a patient’s current mortality status requires matching against external datasets. Thus, the goal is to develop an efficient and secure PPER pipeline for integrating external mortality data with existing EMR systems to ensure accurate and comprehensive mortality information for research purposes.

\noindent\textbf{Datasets} The patient mortality datasets consist of two sub-datasets. Dataset $D_1$ contains cancer patients with deceased information and has the identifiers shown in \Cref{fig:overview}. Dataset $D_2$ contains patients from the EMR and has all of the identifiers as $D_1$. The basic statistics of these datasets are summarized in \Cref{tab:datasets}. Note that both datasets contain additional attributes, but only the identifiers share common information that are useful to match entities.


\begin{table}
    \centering
    \begin{tabular}{cccccc}
        \toprule
        Dataset & \# records & SSN \% & DOB \% & FN \% & LN \% \\
        \midrule
        $D_1$ & 19,274 & 80.2 & 100 & 100 & 99.99 \\
        \midrule
        $D_2$ & 859,725 & 95.3 & 99.95 & 99.99 & 99.99 \\
        \bottomrule
    \end{tabular}
    \caption{Basic statistics of patient mortality datasets.}
    \label{tab:datasets}
\end{table}

\noindent\textbf{Ground truth} Ground truth is vital for developing PPER algorithms, particularly for parameter tuning and performance evaluation. It consists of sampled shared record pairs across two datasets. 
Typically, establishing such a ground truth set requires labor-intensive annotation by human experts. However, in patient mortality datasets, the presence of unique identifiers allows for the straightforward formation of the ground truth set based on these identifiers.
MRNs serve as unique identifiers for patients within healthcare systems, but they are only guaranteed to be unique within a single organization. Hence, as a nationally unique identifier, the SSN is ideal for bridging record pairs in constructing ground truth. Note we do not use SSN directly for PPER because it is not generally available in other healthcare datasets.

\noindent\textbf{Output} The output of PPER is required to be a list of matching record pairs with match scores (0-100\%). Additionally, some of the other identifiers are selectively included for debugging purposes.

\noindent\textbf{Pre-processing} Medical record data vary in the exact representation of the attributes; we therefore perform preprocessing to normalize the SSN and DOB into standard forms across both datasets.
Concretely, We normalize the SSNs to the form \texttt{XXX-XX-XXXX} and transform the DOB representation into a normalized, zero-padded \texttt{MM/DD/YYYY} representation. 
We additionally remove invalid SSNs and perform deduplication on each dataset individually, reducing the size of $D_1$ to 19,274 and $D_2$ to 859,714.



\section{Preliminaries and Related Work}

\subsection{Definitions}
\label{sec:definition}


A formal problem of \textbf{privacy-preserving entity resolution} (PPER) can be framed as a triple $T=(D, M, E)$, where $D = \{D_1 \ldots D_n\}$ represents a collection of $n$ distinct datasets comprising records $r$, each owned by a different data owner $P = \{P_1 \ldots P_n\}$. The encoding or encryption algorithm $E$ is responsible for maintaining the confidentiality of records from each dataset, such that $r$ from each $D$ is transformed into encoded or ciphertext form, denoted as $\enc{r} \in E(D)$, where $\enc{r}$ explicitly indicates that $r$ is in a privacy-preserving representation (i.e., ciphertext in our setting). The match set $M$ contains pairs of records that match between any two datasets among the $n$ parties, that is, $M=\{(\enc{r_i}, \enc{r_j})\,|\,r_i = r_j; \enc{r_i} \in E(D_k), \enc{r_j} \in E(D_m)\}$, where $r_i=r_j$ indicates that $r_i$ and $r_j$ refer to the same entity in the real world. 

Conducting PPER naively requires so-called \textbf{full comparison} $T=\{(r_i, r_j)\,|\,r_i\in D_k, r_j \in D_m\}$ for all the possible pairs between parties, resulting in a total number of comparisons equivalent to the Cartesian product of the sizes of $D_k$ and $D_m$ (i.e. $|T| = |D_k| \times |D_m|$). Since PPER can be computationally heavy, adopting such a quadratic method is impractical. To eliminate unnecessary comparisons, the \textbf{blocking} algorithm is employed to prune $T$ to $T'$ by removing pairs that are unlikely to be part of the solution $M$. Typically, each record $r$ is represented by a set of \textbf{blocking keys} $\beta(r)=(k_1,k_2,\cdots)$, where $\beta(\cdot)$ is the key generation function. Consequently, the \textbf{candidate pairs} are the pairs that share at least one blocking key in common, that is, $T'=\{(r_i, r_j)\,|\,|\beta(r_i) \cap \beta(r_j)|>0, r_i \in D_1, r_j \in D_2\}$.

\subsection{Homomorphic Encryption and CKKS}

\textit{Homomorphic encryption} (HE) allows \ul{the computation to be performed over encrypted data} while preserving the input/output relationship of the function between the plaintext and ciphertext data~\cite{fontaine2007survey,acar2018survey}. In general, an encryption scheme is said to be homomorphic if for some operator $\odot$ over plaintext ($\odot_{\mathcal{M}}$) and ciphertext ($\odot_{\mathcal{C}}$), the encryption function $E$ satisfies: $\forall m_1, m_2 \in \mathcal{M},\ E(m_1 \odot_{\mathcal{M}} m_2) \leftarrow E(m_1) \odot_{\mathcal{C}} E(m_2)$, where $\mathcal{M}$ denotes the message in plaintext and $\mathcal{C}$ denotes the message in ciphertext. $\leftarrow$ means the computation is direct without any decryption in the middle of the process. Therefore, if we let $\odot$ to be $+$, a computation unit can compute $m_1 + m_2$ using the original message from the data owner in encrypted form, that is, $E(m_1)$ and $E(m_2)$, and is able to compute the addition of two encrypted messages without decrypting. Only the data owner with the key can decrypt the message from ciphertext back to plaintext.

\yycomment{rephrase the above text, copied from our KDD paper}

\textit{CKKS (Cheon-Kim-Kim-Song)} \cite{heforarithmeticofapproximatenumbers} is an encryption scheme that falls under the umbrella of homomorphic encryption. It supports approximate arithmetic operations on encrypted real or complex numbers, and can pack multiple numbers into a single ciphertext for efficiency.
CKKS, along with other HE schemes, offers two primary advantages: First, it maintains the accuracy of algorithms compared to their plaintext equivalents while ensuring the confidentiality of the data. Second, it has been rigorously proved to be both semantically \cite{sako2011semantic} and IND-CPA \cite{bellare1997concrete, GOLDWASSER1984270} secure. Therefore, our pipeline, based on AMPPERE, employs the CKKS HE scheme to safeguard data privacy, ensuring that \ul{neither the data owners learn about the data from other parties nor the third parties learn about the individual data or the actual outcome of the computation}.

\subsection{Related Works}
\noindent\textbf{PPER} Initial research has explored using secure one-way hashing and masking algorithms to protect record content \cite{dusserre1995one, quantin1998ensure}, but these methods are limited as they do not account for similar variants of records. Some algorithms use probabilistic data structures, such as Bloom filters \cite{schnell2009privacy, niedermeyer2014cryptanalysis}, but these are vulnerable to frequency attacks and require a balance between data privacy and utility. Other approaches modify string matching algorithms \cite{wang2015efficient, zhu2017efficient} yet their applicability remains restricted to specific algorithms within particular application contexts. With advancements in word/sentence embedding techniques, some methods \cite{scannapieco2007privacy, bonomi2012frequent,li2021improving} transform original records into multi-dimensional embedding vectors that preserve key semantics for similarity comparison and clustering. However, these embeddings are not secure and are susceptible to inversion attacks \cite{song2020information}. Recently, methods utilizing secure multi-party computation or homomorphic encryption-based approaches have been explored \cite{lindell2005secure,ghai2023lessons,yao2021amppere}, offering strong privacy guarantees but are still computationally and resource-intensive.

\noindent\textbf{PPER in healthcare} Despite the numerous proposed PPER methods, their application in the healthcare domain remains limited due to concerns about scalability, explainability, and stability. Datavant~\cite{datavant} is a company that provides commercial PPER services for healthcare datasets, selected by the N3C collaborative. Specifically, their PPER solution~\cite{bernstam2022real, kiernan2022establishing} generates tokens based on the structure of the dataset to represent records and performs matching based on the number of matches of these tokens. These tokens are secured through a two-step private token generation process that first generates master tokens based on SHA-256 \cite{penard2008secure} and then encrypts the tokens to site-specific tokens using AES-128 \cite{rijmen2001advanced}.
Even though this method ensures the identifier is not directly reversible, it has several drawbacks. For instance, because the tokens are hashed, the matching process can only achieve exact matches, making the precision of the matching reliant on the complexity of the token generation rules. Additionally, the same individual-specific tokens are produced across different datasets, which is generally not considered semantically secure. 



\section{Pipeline Implementation and Optimization}
\label{sec:pipeline}
Before diving into our PPER pipeline, we introduce certain baselines upon which our work is built and subsequently compare them.

\noindent\textbf{Naive HE adaptation} This is the most basic version of the PPER pipeline using HE. It simply wraps the ER operation with HE, that is, \textit{Private Set Intersection (PSI)} \cite{chen_psi_acm}. Therefore, such a method finds pairs using a full quadratic comparison over $T$.

\noindent\textbf{Cleartext ER} This is a plaintext version of the ER pipeline. This operates almost in the same way as the privacy-preserving pipeline: It first computes candidate pairs $T'$ with blocking, then computes similarity scores using ER within $T'$.

\begin{figure}[ht!]
    \centering
    \includegraphics[width=0.85\linewidth]{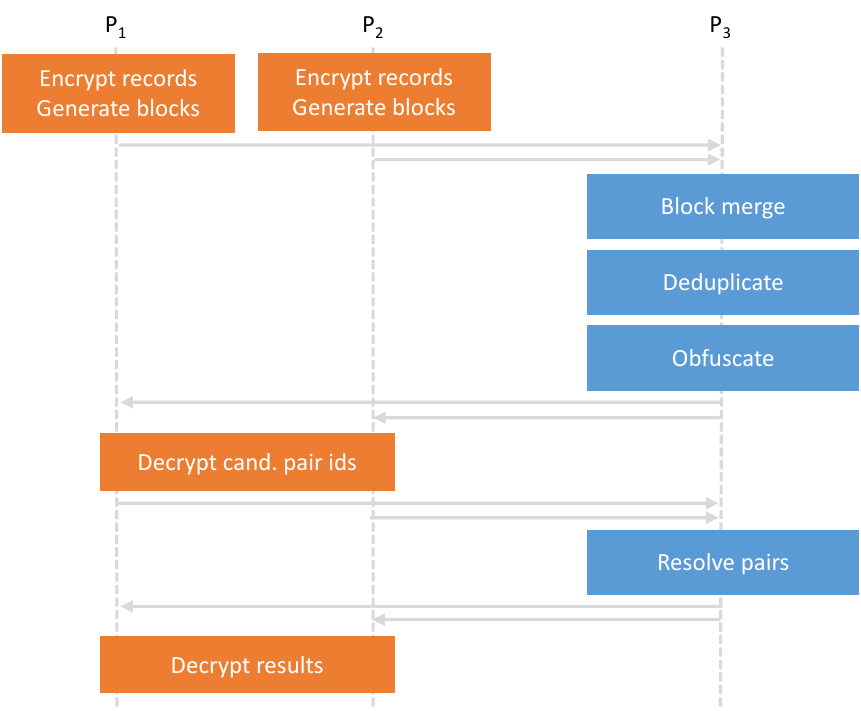}
    \caption{The timing diagram of AMPPERE}
    \label{fig:timing}
\end{figure}

\noindent\textbf{AMPPERE} As introduced before, AMPPERE is an abstract model that defines a cryptographically secure PPER pipeline. Its underlying implementation can be either secure multi-party computation or fully homomorphic encryption. The timing diagram of AMPPERE's execution is in \Cref{fig:timing}. The raw content of the record is homomorphically encrypted in $P_1$ and $P_2$, with blocks generated individually. These encrypted records and blocks are then transmitted to $P_3$ for merging and deduplication using \textit{Private Matrix Operation}. After obfuscation, the entities are resolved using a computationally efficient PSI method called \textit{Vector Rotation}, according to all interactively decrypted candidate pair IDs. The ultimate results are then returned to $P_1$ and $P_2$, where they are collaboratively decrypted.

However, AMPPERE does not scale well to large datasets because of abstract layers for generality. In this work, we chose CKKS for the specific implementation of AMPPERE, primarily because of its efficiency in algorithmic computation and the non-interactive flexibility it offers to the involved parties. We use OpenFHE \cite{openfhe} to implement this scheme. In the subsequent subsections, we illustrate our methodology for constructing the pipeline tailored specifically to medical datasets. We also explore optimization strategies for achieving both maximum accuracy (because CKKS is an approximate arithmetic) and swift execution times. Specifically, in \Cref{sec:chunking}, \Cref{sec:multiprocessing}, and \Cref{sec:simd}, we demonstrate the approach to parallelizing the pipeline from the architecture level, programming level, and algorithm level, respectively. In \Cref{sec:non-intereactive}, we explore the integration of state-of-the-art approximation operators for reduced interactivity. Finally, in \Cref{sec:param-tuning}, we elaborate on the parameter tuning in the context of CKKS.

\subsection{Chunking}
\label{sec:chunking}

\begin{figure}[t!]
    \centering
    \includegraphics[width=.9\linewidth]{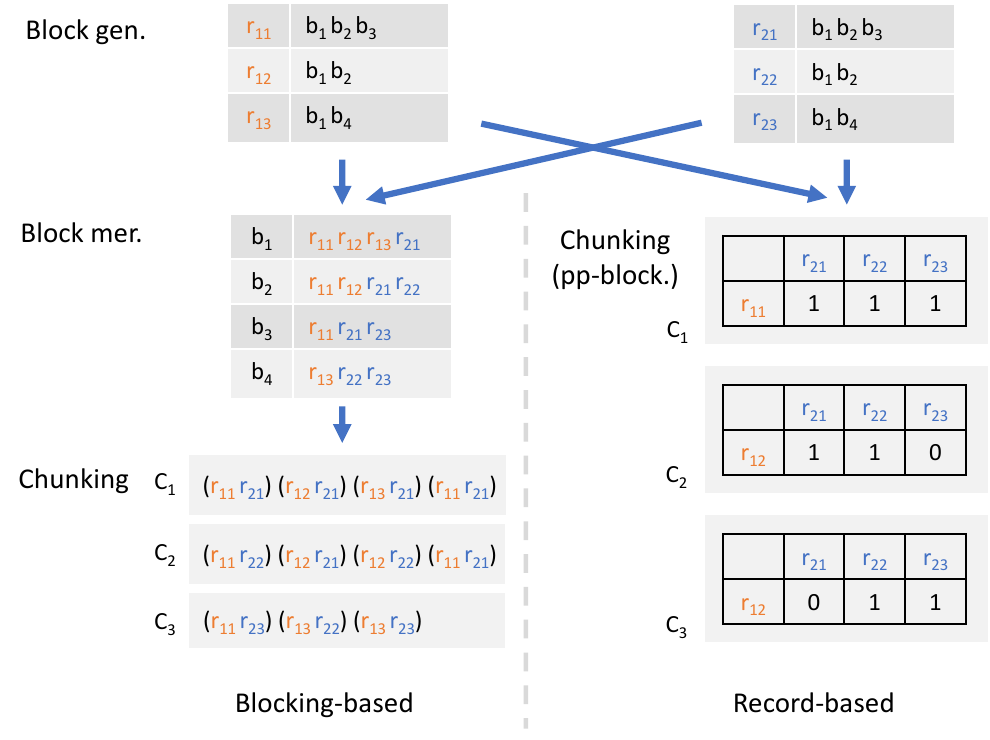}
    \caption{Chunking strategies. Data owners compute blocks independently. In the case of the blocking-based method, these blocks are shared to form candidate pairs, which are subsequently chunked. Whereas for the record-based approach, chunking occurs directly on the records, and candidate pairs are generated via private matrix manipulation within each chunk.}
    \label{fig:chunking}
\end{figure}

Chunking is a common solution for handling large-scale data in ER by breaking down the dataset into smaller ``chunks'' to process them more efficiently with limited computation resources. In the modern ER pipeline, chunking serves an important role in making the original data into manageable units so that they can be consumed and orchestrated by distributed computation frameworks such as MapReduce~\cite{dean2008mapreduce}.
Typically, chunking is based on the output of blocking~\cite{efthymiou2017parallel}. As depicted in the left segment of \Cref{fig:chunking}, after the autonomous generation of blocks in $P_1$ and $P_2$, they need to be shared so that these single-side blocks are merged according to their blocking keys. Subsequently, the candidate pairs for ER are formulated with an optional deduplication step, and these pairs are chunked and distributed to computation units. 

However, at least two issues persist for directly adopting the blocking-based approach. 1) Sharing blocks for merging and deduplication exposes the information of potential entity matches. 2) The overhead of encrypted records and blocks is non-trivial in terms of the process of serialization, deserialization, and data transmission. The blocking-based approach causes a relatively arbitrary distribution, thereby lacking I/O efficiency.

\yycomment{and the total size of the blocks is not consistent}

As a solution, we leverage the record-based approach, as delineated in the right segment of the \Cref{fig:chunking}. The record pairs are first-class citizens and are assigned to chunks along with the associated blocks. In each chunk, blocks are privately merged and deduplicated via AMPPERE's private matrix operation. Consequently, the process of blocking leaks no information and needs no additional I/O-heavy data transmission or serialization/deserialization.

\subsection{Parallel Programming}
\label{sec:multiprocessing}

Modern computers are equipped with multiple CPU cores, and multi-processing/multi-threading are the techniques in which multiple processes/threads are executed concurrently by utilizing multiple CPU cores. We adopt such techniques as programming-level parallelization to utilize the available computation resources fully.

Concretely, decryption is performed in parallel to retrieve the obfuscated pairs, while converting the candidate pair matrix into a candidate pair list is also executed concurrently.
Besides, in more sophisticated scenarios, we orchestrate parallel programming with SIMD (\Cref{sec:simd}). We encrypt and evaluate records as vectors for ease of maintenance and representation consistency, and the concurrent execution of multiple vectors leverages parallel programming. In \Cref{alg:deduplicate}, we eliminate duplicate candidate pairs using a private matrix represented by $M$ encrypted $N$-dimensional vectors. The privacy-preserving update of the matrix is distributed across multiple threads. Moreover, in \Cref{alg:compute_overlaps}, when comparing record similarity, the token comparison using \textit{Vector Rotation (VR)} over encrypted record pair is SIMD-accelerated; meanwhile, multiple VR operations are executed concurrently.

\begin{algorithm}
    \begin{algorithmic}[1]
        \State $cand\_pairs \gets \left[\enc{\operatorname{zeros}(N)}; M\right]$  
        \For{$key \in row\_index$}
            \For{$\enc{row\_id} \in row\_index[key]$}
                \For{$\enc{col\_id} \in col\_index[key]$}
                    \State $cand\_pairs[ \enc{row\_id} ] \gets update($
                    \State $\hspace{2em}cand\_pairs[\enc{row\_id}],$
                    \State $\hspace{2em}\enc{packed\_col\_id},$
                    \State $\hspace{2em}\enc{all\_col\_ids})$
                    \hspace{6.5em}\raisebox{1.5\baselineskip}[0pt][0pt]{$\left.\rule{0pt}{5.2\baselineskip}\right\}\ \mbox{in parallel}$}
                \EndFor
            \EndFor
        \EndFor
    \end{algorithmic}
    \caption{Algorithm for deduplicating using a matrix.}
    \label{alg:deduplicate}
\end{algorithm}



\begin{algorithm}
    \begin{algorithmic}[1]
        \State $overlaps \gets []$
        \For{$(\enc{r_A}, \enc{r_B}) \in enc\_pairs$}
            \State $overlaps.append(\operatorname{VR}(\enc{r_A}, \enc{r_B}))$
            \hspace{3em}\raisebox{0.0\baselineskip}[0pt][0pt]{$\left.\rule{0pt}{1.6\baselineskip}\right\}\ \mbox{in parallel}$}
        \EndFor
    \end{algorithmic}
    \caption{Algorithm for computing the overlaps between records.}
    \label{alg:compute_overlaps}
\end{algorithm}

\subsection{SIMD}
\label{sec:simd}

SIMD stands for Single Instruction Multiple Data, in which a single instruction is applied to multiple data elements at the same time, rather than processing each data element individually. This approach is particularly efficient for tasks in homomorphic encryption evaluation that involve performing the same operation on a large set of data. For example, adding numerous integer pairs sequentially in a loop can be achieved through a single addition operation between two vectors, where the original integer pairs are packed into them respectively.

Despite offering superior efficiency for executing parallel operations on modern computer architectures, integrating SIMD into a homomorphic encryption program poses many challenges. Firstly, not all available operators support parallel execution. Secondly, data must be batched and aligned in advance. Thirdly, certain operations, such as obtaining the vector length, become inaccessible when executed in batch rather than individually.

\jccomment{on review, I found (Thursday or Friday) that AMPPERE already performed vector rotation in a vectorized/SIMD way. I made a first pass at updating this to be in terms of record IDs so that we don't have to update the text or the example too much---let me know what you think.}
Due to the limitations of SIMD, we revamped AMPPERE's representation of the deduplication matrix. Instead of encrypting each record ID individually, the encryption is over the rows of the matrix. Subsequently, we also altered the \textit{Element-wise Equality (EEq)} operation to adapt to the new matrix representation. As an example, when computing the difference between record ID pairs, if each record ID is element-wisely encrypted, the operation is as \Cref{alg:compute_diff} (upper section) where each pair of elements is compared individually in a loop. By contrast, \Cref{alg:compute_diff} (lower section) demonstrates an advanced version where the record IDs are represented by vectors, and encryption is applied to the vectors. With SIMD, the output is obtained in a single operation.

\begin{algorithm}
    \begin{algorithmic}[1]
        \LineComment{With element-wise encryption}
        \State $enc\_a \gets [E(a_1), E(a_2), \dots, E(a_n)]$
        \State $enc\_b \gets [E(b_1), E(b_2), \dots, E(b_n)]$
        \For{$i \gets 1\dots{}n$}
            \State $enc\_output[i] \gets enc\_a[i] - enc\_b[i]$
        \EndFor
    \[\]
    \LineComment{With vector-wise encryption and SIMD}
    \State $\enc{a} \gets E([a_1, a_2, \dots, a_n])$
    \State $\enc{b} \gets E([b_1, b_2, \dots, b_n])$
    \State $\enc{output} \gets \enc{a} - \enc{b}$
    \end{algorithmic}
    \caption{Computing difference between record ID pairs}
    \label{alg:compute_diff}
\end{algorithm}

\subsection{Non-Interactive Comparison}
\label{sec:non-intereactive}

Logical comparisons are fundamental to decision-making in algorithms and control flow structures like loops and conditional statements in programming. AMPPERE utilizes a crucial logical operator $\eeq(a,b)$. This operator returns a Boolean vector that encapsulates the result of element-wise comparisons, indicating whether each corresponding element in $a$ is identical to the respective element in $b$.
AMPPERE provides a workaround for pure algorithmic homomorphic encryption schemes, such as CKKS, that does not support any logical operators: $(a-b) * \frac{-1}{a-b+\xi}+1$, where $\xi$ is a small offset added to prevent dividing by zero. Unfortunately, computing $\frac{-1}{a-b+\xi}$ involves an inverse operation, which is not available in CKKS. As an alternative approach, the original cipher $\enc{x}$ is multiplied by a randomly generated number $\enc{r} \in \{\enc{r_1}, \cdots, \enc{r_n}\}$, pre-generated by $P_1/P_2$, resulting in $\enc{rx}$. This $\enc{rx}$ is then transmitted to $P_1/P_2$ for decryption and inverse computation. The encrypted result $\enc{\frac{1}{rx}}$ is sent back to $P_3$ and ultimately transformed into $\enc{\frac{1}{x}}$ via multiplication with $\enc{r}$, yielding $\enc{\frac{1}{rx}} * \enc{r}$.

In recent years, new solutions have been proposed for computing inverse as well as testing equality without random numbers and interaction between parties. \citet{invCompPaper2019} proposed approximated $\inv$ (inverse) and $\comp$ (comparison, based on $\inv$) solely relying on addition and multiplication, making it possible to be applied in HE. 
The $\comp(a,b)$ function returns 0, 0.5, and 1 for the cases that $a<b$, $a=b$, and $a>b$, respectively. Note that $a$ and $b$ are vectors and $\comp$ is SIMD-accelerated. To use $\comp$, inputs $a$ and $b$ are required to be in a certain fractional range. Yet, employing it to compare record ID equality is not straightforward due to the representation of record IDs as integers. In our implementation, we address this challenge by transforming IDs. Specifically, we assign in-chunk IDs to records, ensuring the chunk size does not exceed the batch size, that is, the in-chunk ID $i \in \{0, 1, \dots, batchSize - 1\}$. Therefore, we rescale $i$ as $\hat{i} = \frac{1}{2} + \frac{i}{batchSize}$, such that $\hat{i} \in \left[\frac{1}{2}, \frac{3}{2}\right)$.
To build $\eeq(a,b)$, the function needs one more step to map the result of $\comp(a,b)$ to Boolean, that is, $0/1 \mapsto 0$ ($a < b$ / $a > b$), and $0.5 \mapsto 1$ ($a = b$). This can be achieved arithmetically by $\eeq(a,b) = 4\comp(a,b)\comp(b,a) = 4\comp(a,b)(1-\comp(a,b))$.



\subsection{Parameter Tuning}
\label{sec:param-tuning}

Tuning parameters in a homomorphic encryption scheme is essential to achieve a balanced trade-off between security, performance, correctness, scalability, and resource utilization. CKKS is no exception. We list all the parameters that are used in \Cref{tab:parameters}. 

Here, we demonstrate the three important parameters that affect the approximation accuracy and running efficiency the most in our implementation. \textit{Batch size} refers to the number of plaintext slots that can be encrypted and processed simultaneously within a single ciphertext. CKKS allows for the packing of multiple plaintext values into one ciphertext, enabling parallel processing of these values, which significantly enhances the efficiency of homomorphic computations. \textit{Scale factor bits} refer to the number of bits used to represent the scaling factor in CKKS. It influences the precision and noise behavior of homomorphic computations. Proper balancing of this parameter is critical for ensuring accurate and efficient encrypted computations. \textit{Number of large digits} refers to the number of digits in key switching, which is a crucial operation for changing encryption keys for ciphers. Choosing a larger digit size simplifies complexity, but it results in a larger key size.
Naturally, we choose our batch size, scale factor bits, and number of large digits to maximize accuracy for candidate pair deduplication via private matrix operation. We find that the SIMD-optimized version of the deduplication algorithm is most accurate with a small batch size, a single large digit, and 50 scale factor bits. We, therefore, choose the minimum batch size needed for our experiments. The chunking experiments require a batch size of at least 100. Since batch size must be a power of two, the minimum batch size is 128.


Another crucial parameter is \textit{multiplicative depth}, which refers to the maximum number of consecutive homomorphic multiplication operations that can be performed on encrypted data before the noise grows too large for successful decryption.
In general, there is no advantage to greater multiplication depth beyond the minimum required depth for the homomorphic circuit to be executed. Therefore, we use the minimum multiplication depth needed in each case.
However, due to multiplicative operations in the $\comp$ function, the non-interactive pipeline requires a prohibitively large multiplication depth. Specifically, in our pipeline, we use a set of $\comp$ parameters $m = 4$ and $(d', d, t) = (5, 5, 6)$ \cite{invCompPaper2019}. With this setting, the minimum required depth is 65. This extreme depth results in keys that use a prohibitive amount of memory.
Instead of increasing the depth, we use \textit{bootstrapping}, a technique that refreshes a ciphertext by reducing its noise, as an alternative. Although bootstrapping enables more homomorphic operations on a fixed pre-set depth without surpassing noise limits, it has significant drawbacks, including high computational overhead, increased latency, resource consumption, and potential precision loss. To balance these factors, we set the multiplication depth to 12 and incorporate bootstrapping to avoid excessive memory requirements.

\begin{table}
    \centering
    \begin{tabular}{l|c}
        \specialrule{.8pt}{0pt}{0pt}
        Parameter & Value \\ \hline
        Multiplicative depth & 2 (12) \\ \hline
        Scale factor bits & 50 \\ \hline
        Batch size & 128 \\ \hline
        Number of large digits & 1 \\ \hline
        First mod size & 60 \\ \hline
        Digit size & 0 \\ \hline
        Security level & HEStd\_128\_classic \\ \hline
        Scaling technique & FLEXIBLEAUTO \\ \hline
        Key switching technique & HYBRID \\ \hline
        Secret key distribution & UNIFORM\_TERNARY \\ \hline
        Interactive bootstrapping compression level & n/a (COMPACT) \\ 
        \specialrule{.8pt}{0pt}{0pt}
    \end{tabular}
    \caption{OpenFHE parameters for CKKS. The parenthesized values are used only in the bootstrapping-enabled non-interactive pipeline.}
    \label{tab:parameters}
    \vspace{-0.8em}
\end{table}
\section{Experiments}

In this section, we conduct extensive experiments. We first detail the settings in \Cref{sec:exp-settings}. The main results are demonstrated in \Cref{sec:exp-main-results}, followed by the ablation studies of the effectiveness of chunking, parallel programming \& SIMD in \Cref{sec:exp-ablation}.

\begin{figure*}[!ht]
    \begin{minipage}{0.32\linewidth}
        \includegraphics[width=.9\linewidth]{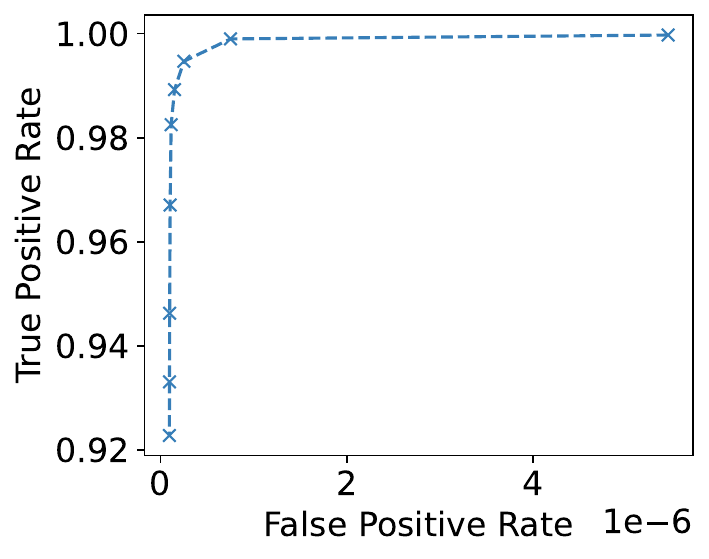}
        \caption{ROC curve for the ER system.}
        \label{fig:exp-roc_curve}
    \end{minipage}%
    \hfill
    \begin{minipage}{0.32\linewidth}
        \includegraphics[width=.9\linewidth]{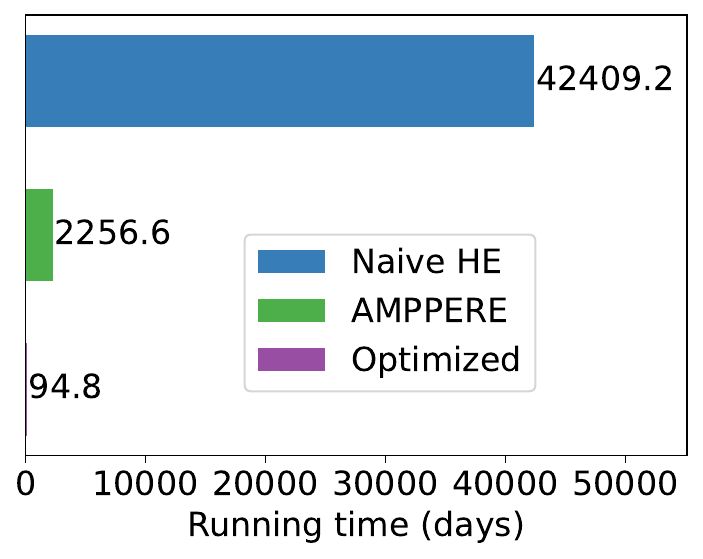}
        \caption{Estimated runtime for our optimized system versus two baselines.}
        \label{fig:exp-main_results_time_comparison}
    \end{minipage}%
    \hfill
    \begin{minipage}{0.32\linewidth}
        \includegraphics[width=.9\linewidth]{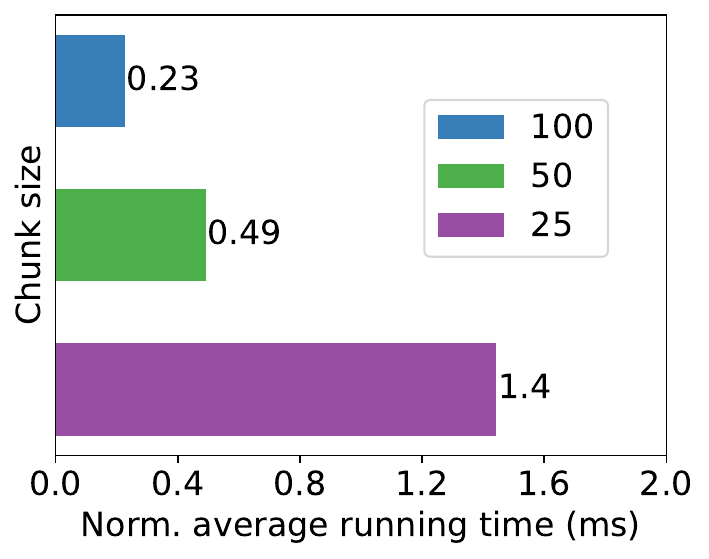}
        \caption{Normalized average time comparison for different chunk sizes.}
        \label{fig:exp-chunking}
    \end{minipage}
\end{figure*}

\begin{figure*}[!h]
    \begin{minipage}{0.32\linewidth}
        \includegraphics[width=.9\linewidth]{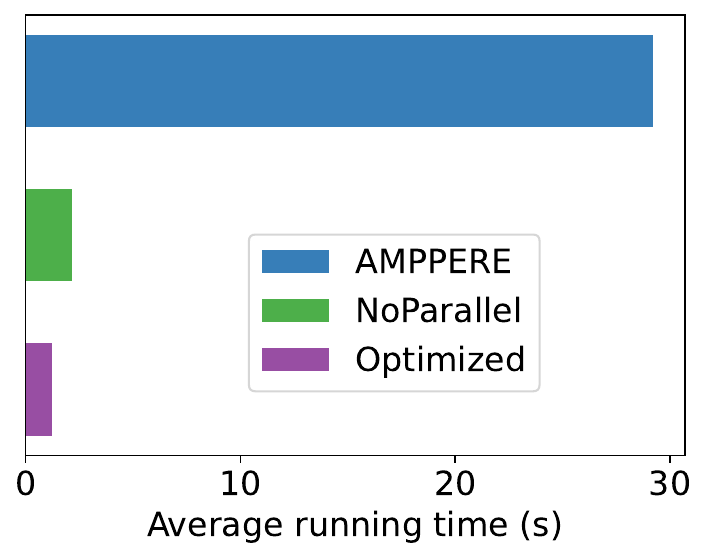}
        \caption{Time comparison for ablation experiments.}
        \label{fig:exp-parallel-programming}
    \end{minipage}%
    \hfill
    \begin{minipage}{0.32\linewidth}
        \includegraphics[width=.9\linewidth]{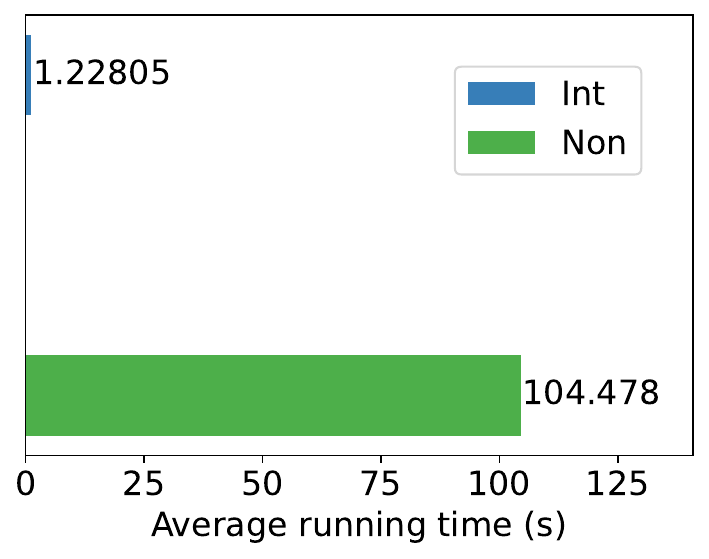}
        \caption{Time comparison for interactive vs. non-interactive pipeline.}
        \label{fig:exp-nointeract_comparison_time}
    \end{minipage}%
    \hfill
    \begin{minipage}{0.32\linewidth}
        \includegraphics[width=.9\linewidth]{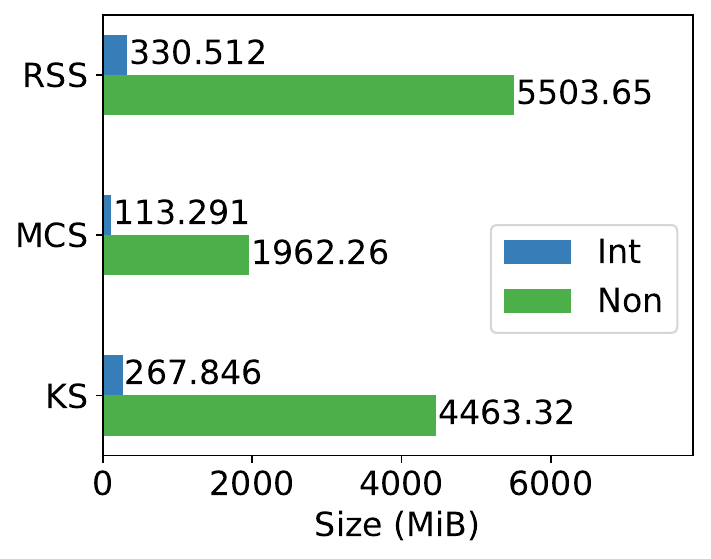}
        \caption{Space comparison for interactive vs. non-interactive pipeline.}
        \label{fig:exp-nointeract_comparison}
    \end{minipage}
\end{figure*}

\subsection{Settings}
\label{sec:exp-settings}
\noindent\textbf{Metrics} We compute \textit{true positive rate} (also called \textit{recall}), \textit{precision}, and \textit{false positive rate} for measuring the ER accuracy. As for the performance of blocking, we use \textit{pairs completeness} ($PC=\tfrac{M \cap T'}{M}$) to measure the percentage of true pairs that are blocked, and \textit{reduction ratio} ($RR = 1 - \tfrac{|T'|}{|T|}$) to measure how well the method reduces the number of candidate pairs. $F\textnormal{-}1$ is employed to harmonically demonstrate the overall performance of blocking.

\noindent\textbf{Environment} The experiments are conducted on a privacy-protected Linux instance with 14G RAM and a 4-core CPU. The pipeline implementation uses OpenFHE version 1.1.2.

\yycomment{spec details}
\jccomment{Yixiang, I updated this paragraph about the sample dataset, but per our discussion we may want to delete the paragraph anyway. Because we've improved our sampling technique, we probably don't need to describe this in detail, but I rewrote the description to describe precisely the updated sampling approach just in case we want it. Feel free to delete it, or to remove everything after ``on randomly sampled subsets.''}



\subsection{Main Results}
\label{sec:exp-main-results}

We report the main results from two perspectives. The first is blocking and ER performance, as well as the cost of running time.

The effectiveness of blocking is represented in \Cref{tab:exp-cleartext_blocking_metrics}. The pair completeness ($PC$) is 100\%, indicating that the blocking algorithm captures all potential pairs. The reduction ratio $RR$ is 99.996\%, meaning the candidate pair set $T'$ is only 0.004\% of the full comparison set $T$. Therefore, applying blocking to this dataset preserves all potential pairs while significantly reducing the pairwise comparisons by more than 99\%.
As for the ER performance, we access it with different thresholds $t$ for the record-matching scores. If the score is greater than $t$, the record pair is identified as the same entity. We show the results in \Cref{fig:exp-roc_curve} using nine evenly spaced thresholds from 0.1 to 0.9. The pipeline achieves a high recall (over 92\%) for the highest threshold, and slightly lower thresholds decrease the precision only negligibly. Moreover, a threshold of around 0.5 achieves nearly perfect recall at approximately 90\% precision and a false positive rate of less than 0.0001\%.
The evaluation above is performed on the cleartext baseline. Both AMPPERE and the optimized pipeline exhibit identical performance because our implementation does not sacrifice any matching performance.

In terms of runtime, we compare our optimized pipeline with the naive HE adaption and AMPPERE. As illustrated in \Cref{fig:exp-main_results_time_comparison}, our optimized system, with chunk size set to 50, operates 24 times faster than AMPPERE and 447 times faster than the naive HE baseline. This notable acceleration underscores the effectiveness of our optimization, leveraging multi-level parallelizations and optimal parameters. In comparison, AMPPERE operates 19 times faster than the naive HE baseline, largely due to its effective blocking and vector rotation for PSI calculation. Note that for our optimized solution, if more computation resources are available, such as with 96 CPU cores, the speed will increase 24 times.

\begin{table}
    \begin{tabular}{rrr}
        \toprule
        Pair Completeness ($PC$) & Reduction Ratio ($RR$) & $F\textnormal{-}score$ \\
        \midrule
        100\% & 99.996\% & 99.998\% \\ 
        \bottomrule
    \end{tabular}
    \caption{Blocking performance for the ER pipeline.}
    \label{tab:exp-cleartext_blocking_metrics}
\end{table}





\subsection{Ablation Studies for Pipeline Optimizations}
\label{sec:exp-ablation}

We report the average time per chunk pair for running time, as it is the fundamental unit into which the data is distributed for computation. This measure allows convenient estimation of time costs with varying computational resources. The average time is computed with the same number of input records for baselines.

\noindent\textbf{Chunking}
As the encrypted dataset cannot fit into memory entirely, we partition it into smaller chunks. This experiment investigates the relationship between running time and chunk size. We examine chunk sizes of 25, 50, and 100 for each dataset. Because the optimized method employs record-based chunking, the potential number of record pairs for a chunk pair equals the square of the original chunk size: $25^2=625$, $50^2=2,500$, and $100^2=10,000$. Consequently, we normalize the running time by the squared chunk size. We present the results in \Cref{fig:exp-chunking}. The findings reveal that doubling the chunk size significantly reduces the normalized running time. This suggests that the increment in chunk size does not have a linear correlation with the normalized running time. The poorer performance observed with smaller chunks may be attributed to more frequent memory swapping, disk I/O, and serialization/deserialization of the cipher data.
As a practical guideline for achieving optimal running speed, the batch size should be set to the maximum possible value based on the available memory size.



\noindent\textbf{Parallel Programming \& SIMD}
%
In this experiment, we assess the impact of parallel programming and SIMD. We conduct two versions for parallel programming: one with parallel programming enabled and one without. For SIMD, due to its changes to the cipher data representation and corresponding functions, it cannot be isolated in the experiments. Consequently, we use AMPPERE as the baseline and create two variants of the optimized pipeline: SIMD-only (NoParallel) and SIMD combined with Parallel Programming (Optimized). The results are presented in \Cref{fig:exp-parallel-programming}. SIMD-only is significantly faster than AMPPERE, highlighting the benefits of processing a batch of data simultaneously. When parallel programming is enabled, the optimized version becomes even faster, saving over 40\% of the time compared to the SIMD-only version.




\noindent\textbf{Non-Interactive Comparison}
%
We first compare the time cost of the interactive and non-interactive pipelines. The non-interactive pipeline (\Cref{fig:exp-nointeract_comparison_time}), although devoid of interaction and pre-generated random ciphers, operates 85 times slower than the interactive pipeline. This discrepancy arises from the relatively high multiplicative depth and the overhead associated with bootstrapping.

Furthermore, we evaluate memory and storage utilization between the interactive and non-interactive pipelines. Specifically, we compare peak memory usage (RSS), average storage usage for each chunk (MCS), and key storage utilized for keys (KS). As demonstrated in \Cref{fig:exp-nointeract_comparison}, the non-interactive pipeline consumes 16.6, 17.3, and 16.6 times more than the corresponding interactive pipeline in each category, respectively. These findings underscore that the non-interactive pipeline necessitates significantly greater memory and storage per chunk.



\section{Conclusion}

In this work, we utilized the CKKS variant within the AMPPERE framework to tackle the challenge of privacy-preserving entity resolution on real-world healthcare datasets. The CKKS homomorphic encryption scheme ensures data privacy by preventing any leakage of confidential information and complies with HIPAA regulations. Furthermore, we comprehensively optimized AMPPERE's pipeline, from data representations to the execution of various operators. These enhancements have significantly improved efficiency without compromising the accuracy of the entity resolution task or the level of privacy protection. This suggests that such approaches are scalable and practical for large-scale applications.


\clearpage
\bibliographystyle{ACM-Reference-Format}
\bibliography{refs}



\end{document}